\documentclass[prb,twocolumn,amssymb,amsmath,nobibnotes]{revtex4}
\usepackage{amsbsy}
\usepackage{bm}
\usepackage{amsmath}
\usepackage{nicefrac}
\usepackage{graphicx,color}
\usepackage{subfigure}
\usepackage{dcolumn}

\begin{document}

\title{Electronic and Transport Properties of Molecular Junctions under a Finite Bias: A Dual Mean Field Approach}

\author{Shuanglong Liu$^1$}
\author{Yuan Ping Feng$^1$}
\author{Chun Zhang$^{1,2}$}
\email{phyzc@nus.edu.sg}
\affiliation{
	$^1$Department of Physics and Graphene research centre, National University of Singapore,
		2 Science Drive 3, Singapore 117542\\
	$^2$Department of Chemistry, National University of Singapore,
	  3 Science Drive 3, Singapore 117543}

\begin{abstract}

We show that when a molecular junction is under an external bias, its properties can not be uniquely determined by the total electron density in the same manner as the density functional theory (DFT) for ground state (GS) properties. In order to correctly incorporate bias-induced nonequilibrium effects, we present a dual mean field (DMF) approach. The key idea is that the total electron density together with the density of current-carrying electrons are sufficient to determine the properties of the system. Two mean fields, one for current-carrying electrons and the other one for equilibrium electrons can then be derived. 
Calculations for a graphene nanoribbon (GNR) junction show that compared with the commonly used \textit{ab initio} transport theory, the DMF approach could significantly reduce the electric current at low biases due to the non-equilibrium corrections to the mean field potential in the scattering region.

\end{abstract}
\maketitle


Since the pioneering work of Aviram and Ratner~\cite{AR_diode}, molecular electronics has attracted a great deal of interests due to its promise for future electronics technology. The central topic in theoretical research of molecular electronics is to understand the quantum electron transport at the molecular level by relating the electric current passing through the molecular junction to its intrinsic electronic properties. The commonly used \textit{ab initio} approach combines the quantum transport theory based on non-equilibrium Green's function (NEGF) techniques and the computational method based on density functional theory (DFT).~\cite{GHPRB,RATNERmethod} 
The approach has been applied to describe the quantum electron transport through various types of molecular junctions.~\cite{Ratnerreview, polymerjcp, zhangcswitch, jeremydiode, Lopeznanolett}, and great success has been achieved in understanding quantum electron transport and also inspiring novel applications in molecular electronics. 

Despite its great success, there are still two problems in the current approach: 1) whether or not the DFT is good enough for molecular junctions under a finite bias is questionable, and 2) when quantitatively compared with experiments, in most cases, theoretically calculated electric current is significantly higher. For some molecular junctions, there might be orders of magnitude differences between experiment and theory. ~\cite{Ratnerreview, polymerjcp, shenleijacs} In this paper, we examine in details the bias-induced noneqiulibrium effects in molecular junctions that are neglected in the current approach and propose a new \textit{ab initio} method to incorporate these nonequilibrium effects.

\begin{figure}
\includegraphics[width=0.5\textwidth]{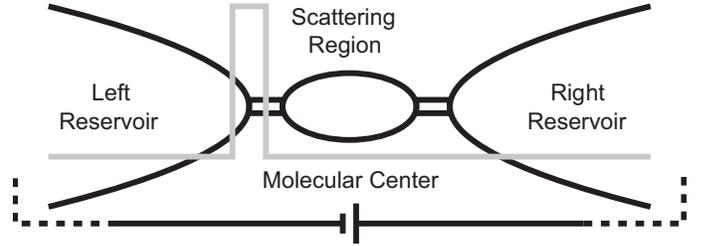}
\caption{\label{fig:model} A molecular scale junction consists of an interacting scattering region and two non-interacting reservoirs. At $t=-\infty$, an infinitely high single-electron barrier potential (the gray line) is applied at the left contact so that no electric current is flowing through. The barrier potential slowly decreases to zero from $t=-\infty$ to $t=0$. The system reaches the desired non-equilibrium steady state at $t=0$.}
\end{figure} 

Molecular junctions connected with two reservoirs can be modeled as open systems with the general structure shown in Fig. 1.
The system is divided into three regions: left, right reservoirs, and the molecular-scale scattering region in the middle. Electrons in two reservoirs are assumed to be in their local equilibrium and can be described by single-electron or mean-field Hamiltonians ($H_L$ and $H_R$ for left and right reservoirs respectively) so that two constant chemical potentials $\mu_l$ and $\mu_r$ can be defined in left and right reservoirs respectively. The bias voltage across the system can then be defined by the difference between chemical potentials in two reservoirs as $V_b=(\mu_l-\mu_r)/e$. Electrons in the scattering region are described by an interacting Hamiltonian $H_S$. The total Hamiltonian for such an open system can be written as $H=H_L+H_R+H_S+H_T \label{Ham}$ where $H_T$ is the tunneling term. The goal of the transport theory is to relate the electronic and transport properties of the steady state of the system to the Hamiltonian and the voltage bias, which is extremely difficult (if not impossible) with the presence of the complicated interaction term $H_S$ in the scattering region.

For simplicity, we set the temperature to be zero throughout the paper. When $\mu_l$ equals $\mu_r$, the whole system is in equilibrium; the DFT is applicable, and then the interaction term $H_S$ can be replaced by the mean-field DFT Hamiltonian. Using NEGF techniques, the mean-field Schr\"{o}dinger equation together with the open-system boundary conditions can be solved,~\cite{GHJTprb,TranSIESTA} and in turn the electronic properties can be worked out. In practice, due to the problems of existing DFT functionals and also the fact that the conductance calculations rely on single-electron orbital, calculations may not be accurate.~\cite{QSYGW,KSHSI,KBSI,CGCOR,KBXC,KSTGW,KSTXC,SGLGW}
When $\mu_l$ is not the same as $\mu_r$, the situation however is different. In this case, the scattering region is driven out of equilibrium, and whether or not its electronic structures in principle can be described by DFT needs to be carefully examined. 

Theoretically, the non-equilibrium steady state of the junction can be achieved by the following adiabatic time-dependent process. At beginning $t=-\infty$, an infinitely high single-electron barrier potential is applied at the left contact as shown in Fig. 1 so that there is no electric current flowing through the system. The two parts separated by the barrier can be denoted as LB (the part on the left of the barrier) and RB (the part on the right of the barrier). The coupling between LB and RB is then gradually turned on by slowly decreasing the barrier potential to zero from $t=-\infty$ to $t=0$. The final state at $t=0$ is the non-equilibrium steady state we desire. For such a time-dependent system, the Runge-Gross (RG) theorem, the foundation of time-dependent DFT,~\cite{TDDFT} claims that the external potential at time $t$, $v_{ext}(\boldsymbol{r},t)$, can be uniquely determined by the time-dependent electron density $\rho (\boldsymbol{r},t)$ together with the initial state $\psi(t=-\infty)$ up to a trivial additive time-dependent function $c(t)$. If initially the system is in a stationary ground state, the initial state $\psi(t=-\infty)$ itself is a functional of initial electron density $\rho (\boldsymbol{r}, t=-\infty)$ according to the first Hohenberg-Kohn (HK) theorem~\cite{HK}, and then the initial-state dependence of the external potential in RG theorem can be eliminated. As a result, at $t=0$, the external potential and in turn the Hamiltonian can be uniquely determined by $\rho (\boldsymbol{r},t=0)$, justifying the commonly used \textit{ab initio} transport theory that combines DFT and NEGF. 

Unfortunately, when the external bias $V_b$ is not zero, initially ($t=-\infty$), the system is in a non-equilibrium state instead of a stationary ground state. The non-equilibrium initial state consists of two separated parts, LB and RB, each of which is in its own equilibrium. From the first HK theorem, we know that the external potential of each part at $t=-\infty$ can be determined by its electron density up to an arbitrary constant. Combining two parts together, the external potential of the whole system at $t=-\infty$ is determined by $\rho (\boldsymbol{r},t=-\infty)$ in whole space up to two independent arbitrary constants $c_l$ (from LB) and $c_r$ (from RB), which can be expressed as $v_{ext} (\boldsymbol{r},t=-\infty) \equiv \widetilde{v}_{ext} [\rho (\boldsymbol{r}, t=-\infty)] + c_l|_{\boldsymbol{r}\in LB} + c_r|_{\boldsymbol{r}\in RB}$. Two constants, $c_l$ and $c_r$, shift the Hamiltonian and chemical potentials in LB and RB, respectively. Only knowing the electron density, $c_l$ and $c_r$ are not determined, then the Hamiltonian of the whole system and also the bias voltage $V_b$ cannot be determined, resulting in a fact that the initial Hamiltonian and in turn the initial state in general can not be uniquely determined by the electron density alone. With given $\widetilde{v}_{ext} [\rho (\boldsymbol{r},t=-\infty)]$, $c_l$ and $c_r$ simply shift $\mu_l$ and $\mu_r$ respectively. Without losing generality, the external potential can also be written as $v_{ext} (\boldsymbol{r},t=-\infty) \equiv \widetilde{\widetilde{v}}_{ext} [\rho (\boldsymbol{r}, t=-\infty)] + \mu_l|_{\boldsymbol{r}\in LB} + \mu_r|_{\boldsymbol{r}\in RB}$. According to the definition of the bias voltage $V_b$,
we have $v_{ext} (\boldsymbol{r},t=-\infty) \equiv \widetilde{\widetilde{v}}_{ext} [\rho (\boldsymbol{r}, t=-\infty)] + eV_b|_{\boldsymbol{r}\in LB} + \mu_r|_{\boldsymbol{r}\in (LB+RB)}$. We see immediately that the initial external potential and in turn the initial state is determined by the initial electron density together with the bias voltage $V_b$ up to a trivial additive constant for the whole system $\mu_r$. Consequently, the initial-state dependence at $t=0$ in RG theorem can be replaced by the voltage dependence, leading to an important theorem that forms the basis of the \textit{ab initio} transport theory: When the system reaches the steady state ($t=0$), its external potential, $V_{ext}(\boldsymbol{r},t=0)$, is uniquely determined by the steady-state electron density together with the bias voltage $V_b$ up to a trivial additive constant. As a direct consequence, the energy of the steady state can be written as a voltage-dependent density functional $E[\rho (\boldsymbol{r}), V_b]$. When $V_b=0$, the functional goes to the commonly used DFT one. 

Next, we show that the external parameter $V_b$ can be determined by intrinsic properties of the system. For this purpose, as shown in Eq. 1, we divide the total electron density of the steady state $\rho_t$ into two parts and name these two parts the equilibrium density $\rho_e$ and non-equilibrium density $\rho_n$, respectively.
\begin{eqnarray}
\rho_t (\boldsymbol{r}) &=&  -i \int_{-\infty}^{\mu_l} G^<(\boldsymbol{r},\epsilon)\mathrm{d}\epsilon = \rho_e + \rho_n \nonumber \\
\rho_e (\boldsymbol{r}) &=&	-i \int_{-\infty}^{\mu_r} G^<(\boldsymbol{r},\epsilon)\mathrm{d}\epsilon \nonumber \\
\rho_n (\boldsymbol{r}) &=&	-i \int_{\mu_r}^{\mu_l} G^<(\boldsymbol{r},\epsilon)\mathrm{d}\epsilon 
\end{eqnarray}
Here we assume $\mu_r < \mu_l$. All physical quantities in Eq. 1 are defined for the steady state at $t=0$. The term $G^<$ is the non-equilibrium lesser Green's function that includes effects of both non-equilibrium distribution and many-body interactions.\cite{GHPRB, RATNERmethod,GHJTprb,TranSIESTA} The non-equilibrium density $\rho_n$ as defined has a physical meaning of the density of current-carrying electrons. Note that according to the definition, $\rho_n$ can also be computed in reservoirs although reservoirs are assumed to be in equilibrium. With the assumption of the mean-field reservoirs and also the lesser Green's function in non-interacting systems~\cite{GHPRB,RATNERmethod,zhancMgO}, it is straitforward to show that given $\rho_e$ and $\rho_n$, the total electron density $\rho_t$ and the voltage bias $V_b$ are determined. Considering the fact that $\rho_e$ and $\rho_n$ can also be determined by $\rho_t$ and $V_b$, we therefore proved the one-to-one correspondence between these two sets of variables.  
Now we have one of the major results of the paper, the foundation of the proposed \textit{ab initio} approach: For molecular junctions under a finite bias, the steady-state properties of the system are uniquely determined by the equilibrium and non-equilibrium electron densities, $\rho_e$ and $\rho_n$, as defined in Eq. 1. The aforementioned voltage dependent energy functional can then be written as $E[\rho_e,\rho_n]$.

We now try to generalize the existing DFT-based \textit{ab initio} transport theory~\cite{GHPRB, RATNERmethod} to include bias-induced nonequilibrium effects. For this purpose, we assume a stationary principle for the non-equilibrium steady state: The variation of the the steady-state energy functional is zero, $\delta E|_{\delta \rho_e, \delta \rho_n}=0$. 
By taking the variations of the energy functional with respect to $\rho_e$ and $\rho_n$ in the scattering region, we obtained two effective mean field equations (Eq. 2), 
\begin{eqnarray}
\left( -\frac{1}{2}\boldsymbol{\nabla}^{2} + v_{ext} + V_H + \frac{\partial e_{xc}}{\partial \rho_{e}} \right) \phi_{j}^{e} = \lambda_{j}^{e} \phi_{j}^{e},\nonumber \\
\left( -\frac{1}{2}\boldsymbol{\nabla}^{2} + v_{ext} + V_H + \frac{\partial e_{xc}}{\partial \rho_{n}} \right) \phi_{j}^{n} = \lambda_{j}^{n} \phi_{j}^{n},
\end{eqnarray} 
where $\phi^{e}$ and $\phi^{n}$ are single-electron orbitals that contribute to $\rho_e$ and $\rho_n$ respectively. In the derivation of above equations, the generalized local density approximation, $E_{xc}=\int e_{xc}\allowbreak [\rho_e, \rho_n] (\boldsymbol{r}) \mathrm{d}\boldsymbol{r}$, is used where $e_{xc}$ is the exchange-correlation energy density of the uniform electron gas. The Hartree potential $V_H$ in the scattering region can be obtained by solving the Poisson equation with correct boundary conditions.~\cite{GHJTprb, TranSIESTA} 
Two coefficients, $\lambda^{e}$ and $\lambda^{n}$, are energies of corresponding orbitals. These two equations have to be solved self-consistently together with the correct open-system boundary conditions, and NEGF techniques are powerful in matching the required boundary conditions. Defining two mean-field exchange-correlation potentials as $V^{e}_{xc}=\frac{\partial e_{xc}}{\partial \rho_{e}}$ and $V^{n}_{xc}=\frac{\partial e_{xc}}{\partial \rho_{n}}$, Eq. 2 suggests that the non-equilibrium electrons or current-carrying electrons experience a different mean-field effective potential from the equilibrium electrons do. We therefore name the proposed method the dual-mean-field (DMF) approach. The DMF equations (Eq. 2) are key results of the paper, which provide the theoretical basis for the investigation of electronic properties of molecular junctions under a finite bias. 

In general, the exchange-correlation energy density $e_{xc}$ can be written as the summation of the exchange part and the correlation part, $e_{xc}=e_x + e_c$. The DMF exchange energy density can be worked out by generalizing the Thomas-Fermi-Dirac (TFD) model~\cite{TFD} to non-equilibrium cases by applying an external bias voltage to the uniform electron gas. By placing the uniform electron gas between two non-interacting reservoirs with different chemical potentials, the exchange energy density (Eq. 3) can be analytically derived (The derivation can be found in supporting information), and then two DMF exchange potentials defined as $V^{e}_{x}=\frac{\partial e_{x}}{\partial \rho_{e}}$ and $V^{n}_{x}=\frac{\partial e_{x}}{\partial \rho_{n}}$ can be computed, 

\begin{eqnarray}
e_x(\rho_{t}, \eta)	= \frac{1}{4}\left(1+\eta\right)^{\nicefrac{4}{3}} \bigg[ -\left(1-\tilde{\eta}^{2}\right)^{2}ln\left(1+\tilde{\eta}\right) \nonumber \\
	+ \tilde{\eta}^{4}ln\left(\tilde{\eta}\right) + \tilde{\eta}^{4} + \tilde{\eta}^{3} - \frac{1}{2}\tilde{\eta}^{2} + \tilde{\eta} + \frac{3}{2} \bigg]e_{x}^{TFD}(\rho_{t}) ,
\end{eqnarray}
where $\eta(\boldsymbol{r}) = \rho_n (\boldsymbol{r})/ \rho_t (\boldsymbol{r})$ which is called the non-equilibrium index in this paper, and $\tilde{\eta} = \left(\frac{1-\eta}{1+\eta}\right)^{\nicefrac{1}{3}}$. According to the definition, $\eta$ takes the value between 0 and 1, which measures the extent of the nonequilibrium. When $\eta = 0$, the exchange density reduces to the equilibrium TFD one $e_{x}^{TFD}(\rho_{t})=-\frac{1}{4\pi^{3}} \left(3\pi^{2}\rho_{t}\right)^{\nicefrac{4}{3}}$.   
The DMF correlation energy density is however challenging. In this paper, we set the correlation functional to be the same as the DFT one, which may not be a bad approximation for weakly correlated systems under low biases. 

\begin{figure}
\includegraphics[width=0.5\textwidth]{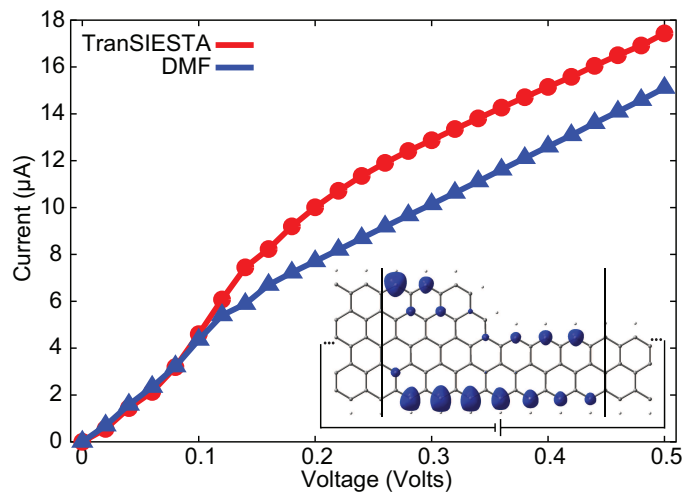}
\caption{\label{fig:iv} I-V curves for a GNR junction calculated from both DMF approach and the commonly used \textit{ab initio} transport theory via the software TranSIESTA. Inset: The atomic structure of the GNR junction and the iso-surface of the difference between the exchange energy potentials of current-carrying electrons calculated from DMF approach and TranSIESTA,$\delta V = V_{x}^n - V_{x}^{TranSIESTA}$. The exchange potentials were calculated under 0.2 V. The iso-surface value is 15 $meV$.}
\end{figure} 

We now are ready to apply the DMF approach to real molecular scale junctions. As a case study, we choose a junction made of two zigzag graphene nanoribbons (Z-GNRs) with different widths as shown in the inset of Fig. 2. The GNRs have been regarded as one of the most promising building blocks for graphene-based electronic devices.~\cite{NetoReview} 
Since the long-range magnetic order may not be stable under a finite temperature for such a one-dimensional system, we follow previous studies to set the total spin of the system zero.~\cite{LFGNR, GNRsymmetry} For Z-GNR based junctions under low biases, It has been well known that the current flows through edge states. 
We have implemented the DMF approach into the SIESTA computational package.~\cite{Siesta} For comparison, we performed calculations with both the DMF approach and the commonly used DFT based transport method via the function TranSIESTA built in SIESTA~\cite{TranSIESTA} (More computational details can be found in supporting information).

In Fig. 2, I-V curves from DMF and TranSIESTA calculations for the GNR junction are presented.
When bias is small ($<$ 0.1 V), the DMF approach essentially reproduces the TranSIESTA results. Starting from 0.1 V, significant deviations between two approaches occur, and the current from DMF calculation is always lower than that calculated from TranSIESTA. To understand this, we plot in the inset of Fig. 2 the iso-surface of the difference between the DMF non-equilibrium exchange potential ($V_{x}^n$) and the DFT exchange potential calculated from TranSIESTA. The potentials were calculated for the bias voltage 0.2 V. We see that the exchange potential increases significantly at edges which are places the electric current flows through. For other parts of the system, the non-equilibrium correction to the potential is not that important. The increase of the exchange potential leads to a higher scattering barrier in the scattering region, and in turn, decreases the current as we see in Fig. 2.

\begin{figure}
\includegraphics[width=0.5\textwidth]{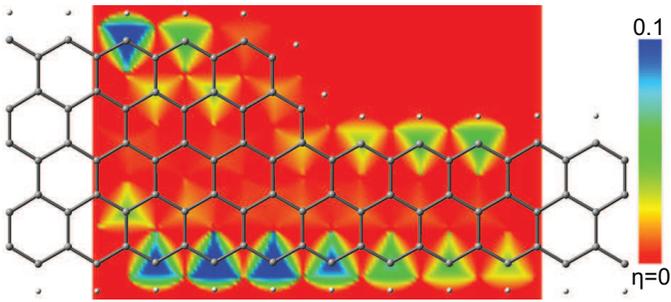}
\caption{\label{fig:ni} Non-equilibrium index $\eta (\boldsymbol{r})$ in the scattering region of the GNR junction. The color map was plotted in the plane 2.2 $\AA$  above the GNR. The system goes from local equilibrium to non-equilibrium when the color changes from red to blue.}
\end{figure} 

The non-equilibrium index in the DMF approach, $\eta (\boldsymbol{r})$, provides detailed spacial information for the non-equilibrium steady state of the system. To demonstrate this, in Fig. 3, we plot the color contour map of $\eta (\boldsymbol{r})$ in the scattering region of the GNR junction under 0.2 V.
From the figure, we can see that in the scattering region, the extent of the non-equilibrium at different places are quite different: Edges are far away from equilibrium while electrons in the center of ribbons are still approximately in local equilibrium.

In conclusion, we propose a DMF approach to describe electronic and transport properties of molecular-scale junctions under a finite bias. We show that two electron densities, $\rho_e$ and $\rho_n$, are needed to uniquely determine the properties of the steady state of non-equilibrium molecular junctions. Subsequently, two mean fields, one for current-carrying electrons and the other one for equilibrium electrons, can be derived. The transport properties can then be calculated from the mean-field potential that the current-carrying electrons experience. The case study for a GNR junction shows that the DMF approach could lead to significantly lower electric current than the commonly used transport theory. 
For molecular junctions that have localized molecular orbitals in the scattering region, the non-equilibrium corrections to the mean-field potential in the DMF approach will cause significant variations of these localized orbitals and lead to more profound changes in transport properties, which will be discussed in our future studies.


We acknowledge the support from Ministry of Education (Singapore) and NUS academic research grants (R-144-000-325-112 and R144-000-298-112). Computations were performed
at the Graphene Research Centre and Centre for Computational Science and Engineering at NUS. See Supplementary Material Document No.xxxxx for the generalized TFD model and computational details. For information on Supplementary Material, see http://www.aip.org/pubservs/epaps.html. The computational codes for all calculations are available online or from the author.




\end{document}